\pdfoutput=1
\documentclass[conference]{IEEEtran}
\IEEEoverridecommandlockouts
\usepackage{cite}
\usepackage{amsmath,amssymb,amsfonts}
\usepackage{algorithmic}
\usepackage{graphicx}
\usepackage{textcomp}
\usepackage{xcolor}
\def\BibTeX{{\rm B\kern-.05em{\sc i\kern-.025em b}\kern-.08em
    T\kern-.1667em\lower.7ex\hbox{E}\kern-.125emX}}
\begin{document}

\title{A case study on profiling of an EEG-based brain decoding interface on Cloud and Edge servers
}

\author{
Alexandra Samsonova\inst{1}
\and
Barry J. Devereux\inst{1}
\and
Georgios Karakonstantis\inst{1}
\and
Lev Mukhanov\inst{1}
}

\author{\IEEEauthorblockN{1\textsuperscript{st} Alexandra Samsonova}
\IEEEauthorblockA{\textit{EEECS} \\
\textit{Queen's University Belfast}\\
Belfast, UK \\
sacha.samsonova@gmail.com}
\and
\IEEEauthorblockN{2\textsuperscript{nd} Barry J. Devereux}
\IEEEauthorblockA{\textit{EEECS} \\
\textit{Queen's University Belfast}\\
Belfast, UK \\
B.Devereux@qub.ac.uk}
\and
\IEEEauthorblockN{3\textsuperscript{rd} Georgios Karakonstantis}
\IEEEauthorblockA{\textit{EEECS} \\
\textit{Queen's University Belfast}\\
Belfast, UK \\
G.Karakonstantis@qub.ac.uk}
\and
\IEEEauthorblockN{4\textsuperscript{th} Lev Mukhanov}
\IEEEauthorblockA{\textit{EEECS} \\
\textit{Queen's University Belfast}\\
Belfast, UK \\
L.Mukhanov@qub.ac.uk}
}

\maketitle

\begin{abstract}
Brain-Computer Interfaces (BCIs) enable converting the brain electrical activity of an interface user to the user commands. 
BCI research studies demonstrated encouraging results in different areas such as neurorehabilitation, control of artificial limbs, control of computer environments, communication and detection of diseases. Most of BCIs use scalp-electroencephalography (EEG), which is a non-invasive method to capture the brain activity.
Although EEG monitoring devices are available in the market, these devices are generally lab-oriented and expensive. Day-to-day use of BCIs is impractical at this time due to the complex techniques required for data preprocessing and signal analysis. This implies that BCI technologies should be improved to facilitate its widespread adoption in Cloud and Edge datacenters. 

This paper presents a case study on profiling the accuracy and performance of a brain-computer interface which runs on typical Cloud and Edge servers. In particular, we investigate how the accuracy and execution time of the preprocessing phase, i.e. the brain signal filtering phase, of a brain-computer interface varies when processing static and live streaming data obtained in real time BCI devices. We identify the optimal size of the packets for sampling brain signals which provides the best trade-off between the accuracy and performance. Finally, we discuss the pros and cons of using typical Cloud and Edge servers to perform the BCI filtering phase. 
\end{abstract}

\section{Introduction}
The discovery in 1924 by Hans Berger of the electrical activity of human brains and possibility of recording it with EEG (electroencephalography) enabled the development of systems for brain-computer interfaces (BCIs). BCI was mentioned for the first time in a paper of Jacques J. Vidal in the 1970s, who used visual evoked-potentials to establish a direct brain-computer communication \cite{doi:10.1146/annurev.bb.02.060173.001105}. Since then, research in understanding the brain functions made possible advances in the development and wider application of BCIs. The state-of-the-art BCIs use electrical potentials which can be measured in a non-invasive way by placing electrodes on the scalp (scalp-EEG), placing electrodes on the exposed surface of brains (ECoG: electrocorticography) or implanted directly into brains. EEG is the most convenient technique since it does not require any surgical procedure and provides a high accuracy\cite{bekaert:hal-00521052}. 

Many BCI-based studies\cite{Roman-Gonzalez2012} are aimed at helping people with various movement disorders, such as locked-in syndrome, by restoring their ability to communicate with others. Furthermore, with the advancement of computer technologies, the use of EEG data has been expanded to provide a means to control of various objects: for example, a cursor on a monitor, a wheelchair \cite{1492493}, or a robotic-hand \cite{HORTAL2015181}. Recently, the possibility of applying BCIs for neurorehabilitation,  which aims to aid recovery from nervous system injuries, has been discovered \cite{MCFARLAND2017194}. In addition, multimedia, communication, and entertainment companies have expressed interest in using BCI as a new way for people to interact with computing devices. Nevertheless, several challenges have been reported in previous BCI studies.

The main challenge is that the vast majority of the BCI studies involve small group of participants (typically less than 25) \cite{SKOLA201859}. This is mostly due to the fact that a BCI interface should be trained for a specific person, and the training procedure, as well as the signal processing and preprocessing phases, requires a lot of computing resources and time. These issues are particularly problematic for real time mobile BCI where a high performance and energy efficiency are required. 
To the best of our knowledge, none of the existing systems is applicable for daily use by disabled persons or for more common applications, e.g. multimedia systems\cite{SKOLA201859}. The BCI technologies are not advanced enough to be tested outside of laboratories. Apart from this, recent studies use only publicly available prerecorded data, which does not guarantee that the proposed BCIs will have the same performance in real systems \cite{MCFARLAND2017194}.
Nevertheless, EEG-based BCI is promising technology that will revolutionize the communication between humans and computers, even though current EEG-based brain decoding interfaces suffer from technical limitations that need to be addressed. 

Many computing services run in Cloud datacenters these days; Cloud service market is expected to reach \$$927$ billion by 2027 \cite{cloudmarket}. Moreover, it is projected that the number of intelligent Internet-connected devices will soon reach tens of billions. All these devices use Internet and Cloud datacenters to transfer and store the data. As a result, the size of data transferred through Internet will exceed 24.3 exabytes soon \cite{ciscoreport}.
To address this challenge, a concept of Edge computing was recently introduced. The main idea behind this concept is to place computing devices close to the sources of data\cite{ibmreport}, which enables the user to significantly reduce the latency of applications allocating data in Cloud and lower the pressure on the network. BCI can be implemented as a Cloud or Edge service, which, we believe, will facilitate the usage of such interfaces by the users. However, the BCI response time should be very small to provide a high quality service for a BCI device. The goal of our study is to investigate the performance and accuracy of BCI when running on typical Cloud and Edge servers.

In this paper, we present a preliminary study on the accuracy and performance of a brain decoding interface, with a focus on practical issues in BCI development, such as hardware suitability and efficient real time data processing. In our study, we focus on the preprocessing stage of the BCI interface, i.e. the filtering phase for EEG signals which implements a band-pass filter. We investigate how the accuracy and performance of this phase changes with the sampling rate and duration of the signal processing period. Then we test the performance of the filtering phase on typical Cloud and Edge servers. We make conclusions about efficiency of the servers for running the filtering phase and the most optimal configuration of the signal sampling rate which provides the best trade-off between the accuracy and performance. 

The contribution of this paper can be summarized as follows:
\begin{itemize}
    \item We present the results of our preliminary study on profiling the accuracy and performance of a brain-computer interface. In particular, we demonstrate the profiling results for the preprocessing phase, or the brain signal filtering phase, of this interface which has a significant impact on the accuracy of the entire BCI pipeline.  
    \item We demonstrate how the accuracy and performance of this phase varies when processing static data and live steaming data, used in real time BCI devices. We identify the optimal size of packets for real time processing which provides the best trade-off between the accuracy and performance.   
    \item Finally, we profile the filtering process on typical Cloud and Edge servers. We demonstrate the pros and cons of each server when running the filtering phase of the brain-computer interface.
\end{itemize}

The rest of the paper is organized as follows. Section  II presents the background and previous studies. Section III introduces the experimental  setup. Section IV demonstrates  the  results of our experimental campaign. Finally, conclusions are drawn in Section V.

\section{Background and previous work}
\begin{figure}[htb]
        \centering
        \includegraphics[height=2.5in,keepaspectratio]{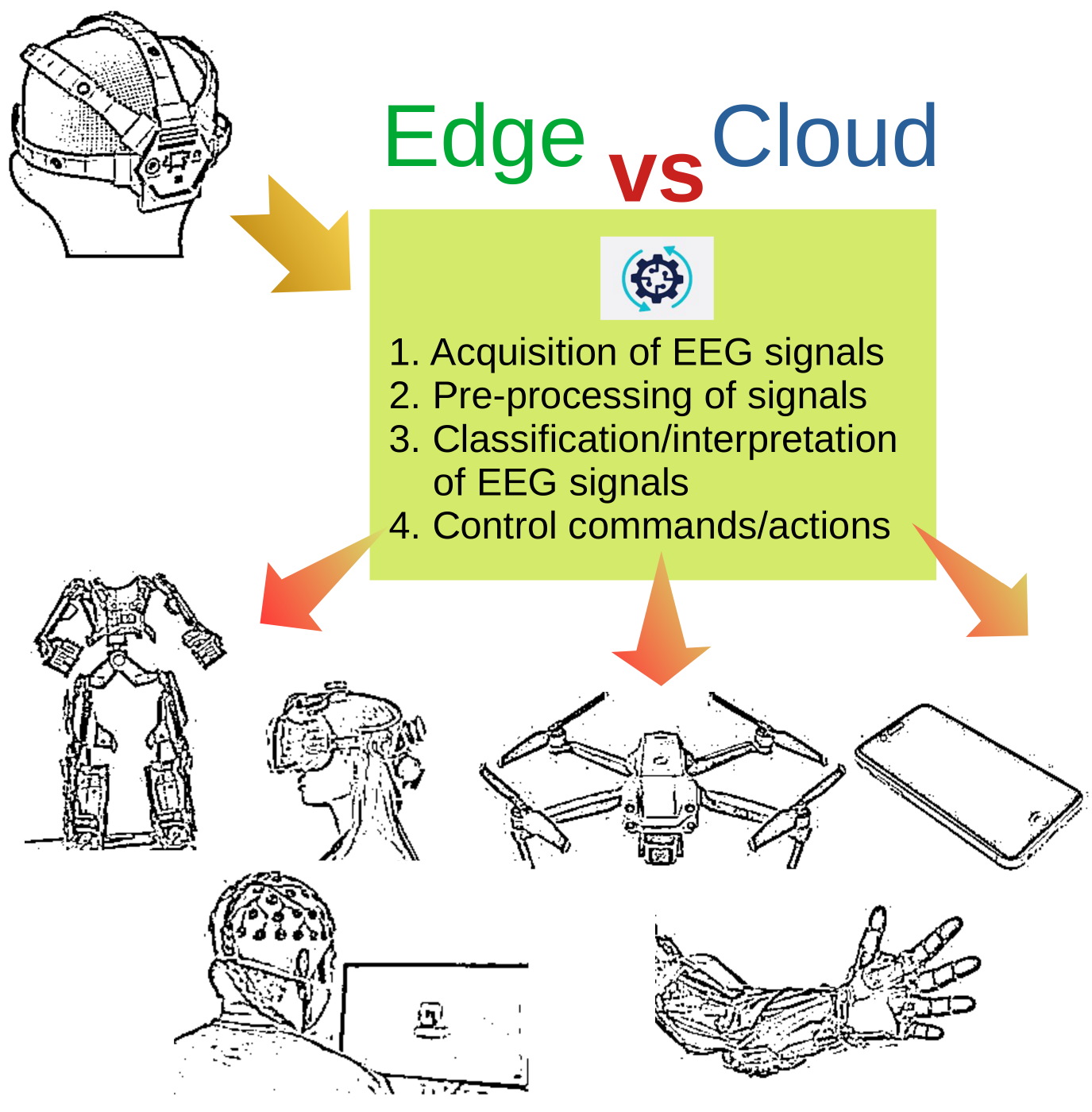}
        \caption{Overview of Brain-Computer Interfaces.}
        \vspace{-0.8cm}
        \label{fig:bci_overvirew}
\end{figure}
\subsection{EEG-based Brain Decoding Interfaces}
EEG-based BCI systems traditionally process brain signals in 4 steps \cite{MCFARLAND2017194} (see Figure \ref{fig:bci_overvirew}). The first step is to receive the EEG signals (the acquisition phase). The second step is processing of the signal to restrain them to a desired frequency (the preprocessing phase). This step is required to remove the noise from the brain signals induced by movement of the muscles, movement of the eyes, eyelids and ambient electrical interference \cite{bekaert:hal-00521052}. The preprocessed signals then are analyzed to predict a user command at the interpretation phase. Finally, based on the predicted command a specific action is performed. This action may include controlling a game, drone, an augmented reality application and even exoskeleton. 

\subsection{Data acquisition and preprocessing}
EEG sensors are widely available in different sizes and at different prices. Some of the EEG based BCI were engineered with specially created devices: Gonzales et al. \cite{Roman-Gonzalez2012} used international system of positioning 10/20 to fix the electrodes and used the Procomp Infinity signal amplifier. 
Škola et al. \cite{SKOLA201859} reported the use of a lightweight wireless EEG device Enobio 32. Qin et al. used an Emotive headset (Emotive Epoc) to implement a cheap and mobile BCI device\cite{QIN2018115}. Biosemi ActiveTwo is an example of a laboratory device, however it is expensive and cannot be used outside laboratories\cite{active2}. 

The brain signals reflect brain activity which has different rhythms within specific frequency bands. The preprocessing steps need to be implemented to remove irrelevant noise in the acquired data. Nevertheless, the performance of the pre-processing phase depends on the EEG signal quality and position of the electrodes. For instance, a study be Radüntz et al. \cite{10.3389/fphys.2018.00098} reported that a large proportion of artefacts are created by a poor fit of the headset, while gel-based devices outputted the best signal-to noise ratio (SNR) value. The research literature provides a wide array of EEG preprocessing techniques. Among the noise filters, the Laplacian filter has been first used by McFarland et al. in 1997\cite{MCFARLAND1997386}. Digital band-pass filters can be applied over the signal to isolate the typical neurocognitively relevant range of 3-40Hz\cite{Roman-Gonzalez2012}. A high-pass filter has been applied by Qin et al. \cite{QIN2018115} in order to focus only on the fluctuations reflecting the repetitive flickering. Škola et al. \cite{SKOLA201859} have filtered the signals in a range of 8-30Hz using a 5-th order Buterworth filter. These filtering methods are common for offline EEG processing but not real time. More sophisticated spatial filtering has also been applied\cite{bekaert:hal-00521052}. However, such filters need to be adjusted to the requirements of a particular study and to the brain activity of a specific person \cite{bekaert:hal-00521052}. Among these kinds of filters, Principal Component Analysis (PCA) and Independent Component Analysis (ICA) have been used to separate the noise and extract singlas. Recent studies have also investigated convolutional neural networks, common spatial patterns and source imagery methods that can be applied for filtering\cite{MCFARLAND2017194}. However, such filtering techniques should be adjusted for a particular BCI system. Moreover, Islam et al. \cite{ISLAM2016287} raised concerns regarding efficiency of these techniques due to the lack of completely uncontaminated data.

\subsection{Feature extraction and the interpretation phase}
To build an efficient BCI, it is important to quickly associate EEG signals with user commands\cite{QIN2018115}. This association is implemented using a classifier, which is usually based on a Machine Learning method \cite{SKOLA201859}. Two characteristics define such a classifier: auto-regressive adaptive parameters (AAR) and Spectral Energy in Mu and Beta Band (PST). A previous study implemented the classifier based on a neural network and a linear discriminant analysis\cite{SKOLA201859}. In particular, authors used shrinkage Linear Discrimination Analysis (sLDA) with a regular covariance matrix to implement BCI training for a virtual reality framework\cite{SKOLA201859}. Another study used Bayesian linear discriminant analysis (BDA) to train the classifier \cite{4412807}. Ansari et al. \cite{ANSARI2019274} compares the efficiency of different classifiers, such as SVM, Random Forest, ELM, OS-ELM, K-Nearest Neighbors, Decision Tree and Multilayer perception, that can be applied in BCIs. For real time processing, Gonzalez et al. \cite{Roman-Gonzalez2012} proposed to compress the filtered signals without losing  essential information.

Many studies implemented Machine Learning classification of the brain signals for EEG-based BCIs using MATLAB. However, other software packages were also applied for the classification, such as Thought Translation Device (TTD) interface, Openvibe application, BrainVision Recorder and Brain vision Analyzer \cite{ANSARI2019274}.

Overall, we may conclude that BCIs require significant computing resources to process and classify brain signals. The computing platform used by BCIs is important for quick processing of brain signals at runtime. Thus, the following questions arise: i) What hardware platforms should be used to run BCIs? ii) How various BCI configuration parameters affect the accuracy and processing performance? iii) How efficient server-grade hardware platforms used in Cloud for processing signals? iv) Can brain signals be quickly processed on energy efficient Edge servers? 

In this paper, we present the results of our preliminary study on profiling the most important phase in BCIs, i.e. the preprocessing phase, which has a significant impact on the accuracy and performance of any brain-computer interface. 
\section{Experimental setup}
\begin{figure}[htb]
        \centering
        \includegraphics[height=2in,keepaspectratio]{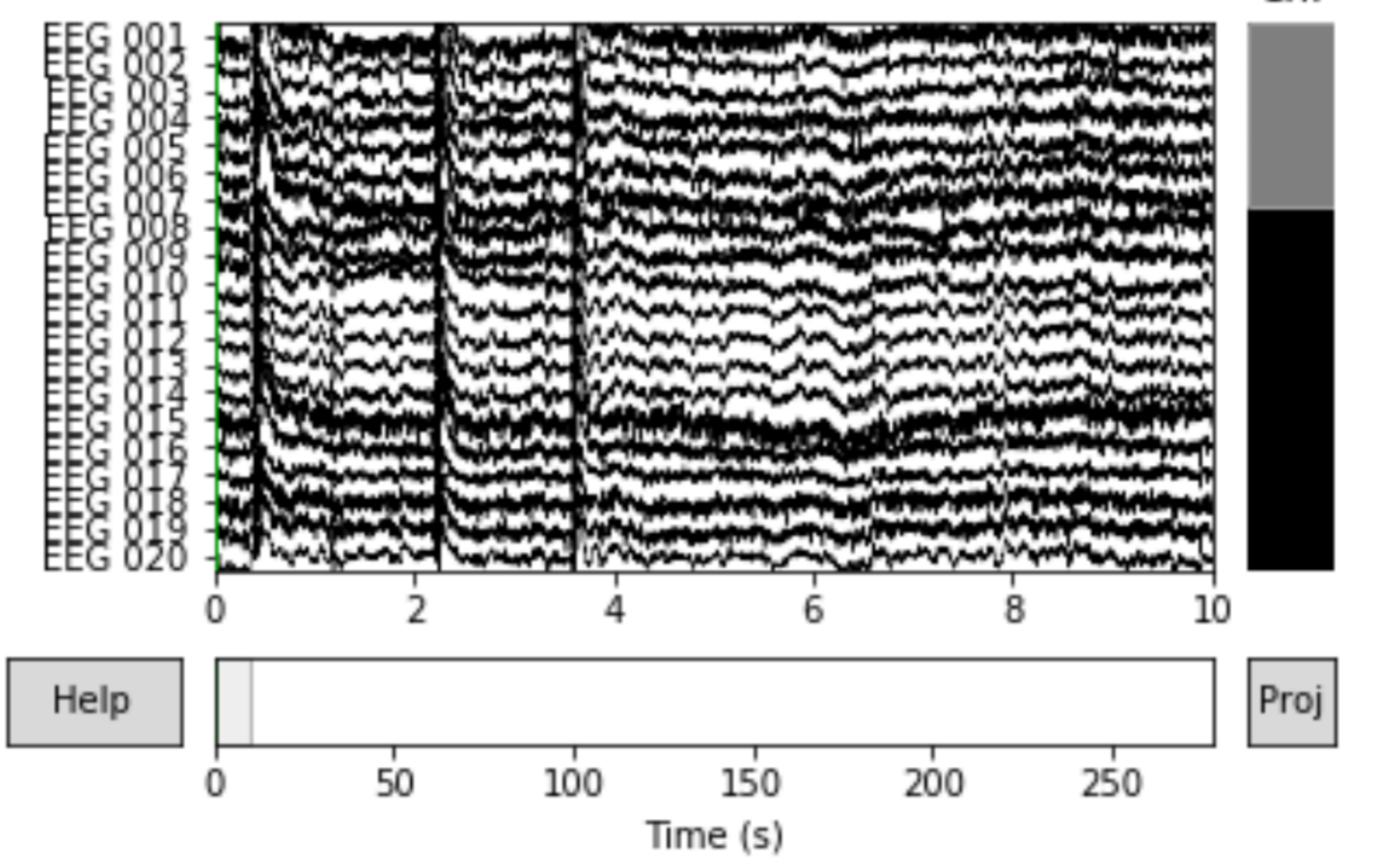}
        \vspace{-0.4cm}
        \caption{First 20 channels of EEG raw signals, the period of 10 seconds.}
        \vspace{-0.4cm}
        \label{fig:eeg_example}
\end{figure}

\subsection{The BCI library and data}
In our study, we use the EEG data extracted from datasets of the MNE library, an open-source Python-based framework for exploring brain activity data (such as MEG, EEG, sEEG, ECoG) \cite{GRAMFORT2014446,10.3389/fnins.2013.00267}. The core data structure of an EEG continuous data set is a 2D NumPy array, combined with an “Info” object, containing details about the data set such as: the sampling rate, digitized points, sensors, recording details and data channels. MNE provides a set of functions configured to take the data and its information as parameters, which are used to visualise and pre-process the EEG data.

The provided EEG data set is composed of 59 channels and 166800 samples, corresponding to the columns and the rows of the 2D array. The duration of a signal in each channel is 277.7 seconds, i.e. 4.62 minutes. The raw EEG data is illustrated in Figure \ref{fig:eeg_example}.

To filter the data,  we use a zero-phase Finite Impulse Response filter ( FIR filter, \textit{firwin}), since FIR filters are: i) easier to control; ii) stable and iii) have a well-defined passband\cite{WIDMANN201534}. To eliminate recorded signals that do not belong to the brain electrical activity, the sampling frequency is set between 2 and 30 Hz.

\subsection{Hardware platform}
In our study, we use a typical Intel-based dual-socket server. Each socket hosts an Intel Xeon E5-2650 (Sandy Bridge) processor featuring an integrated memory controller (iMC) to control the DRAM devices attached to the socket, specifically four 8 GB DDR3 DIMMs at 1600 MHz. Each processor contains 8 cores. In our experimental study, we use the maximum processor frequency, which is 2.9 GHz.

As an edge server, we use a server featuring the AppliedMicro/Ampere X-Gene 2 ARMv8 processor (8 cores in total) \cite{singh2014appliedmicro}. All memory operations in X-Gene 2 are handled by 4 MCUs. Each MCU can be populated with up to 2 DDR3 DIMMs running at 1600 MHz. In our setup, we use 4 DDR3 8GB DIMMs, one for each MCU. Similar to the experiments with the Intel-based server, we set the maximum processor frequency, which is 2.4 GHz.

Finally, for the first experiments of our study we use a desktop-grade system which uses the 4-th generation Intel Core i7 processor(4 cores) and 32GB of DDR3L memory operating at 1600 MHz. We set the maximum frequency for this processor to 2.80 GHz. 
\section{Experiments}
In our study, we use the pre-recorded data for processing the brain signals and simulate real time brain signals by streaming this data. In our initial experiments, we divide a 2D array,  which represents the sampled EEG signals, in chunks, or packets, of the same size. We split the signals into 417 packets and thus each packet contains 400 data points. 
By dividing the brain signals into packets, we simulate a data stream from a real time EEG headset. Most of the previous BCI studies have implemented filters to preprocess EEG data by removing electrical signals induced by activities other than those aimed at driving a BCI. Such filters significantly reduce the noise level in the brain signals.

In our study, we use the filtering function from the MNE library to process the raw EEG data. We filter the brain signal in the range from 2Hz to 30Hz to eliminate signal noises that do not belong to the brain electrical activity.
Thus, we use the MNE band-pass filter as a zero-phase Finite Impulse Response (FIR) filter, which has been highly recommended for the electrophysiological data analysis \cite{WIDMANN201534}.
Note that to raise the filtering execution time to be enough for profiling, we increase the size of the EEG data by copying the same fragment three times.

\begin{figure}[hbt!]
        \begin{minipage}[t]{0.48\linewidth}
                \includegraphics[width=\columnwidth,keepaspectratio]{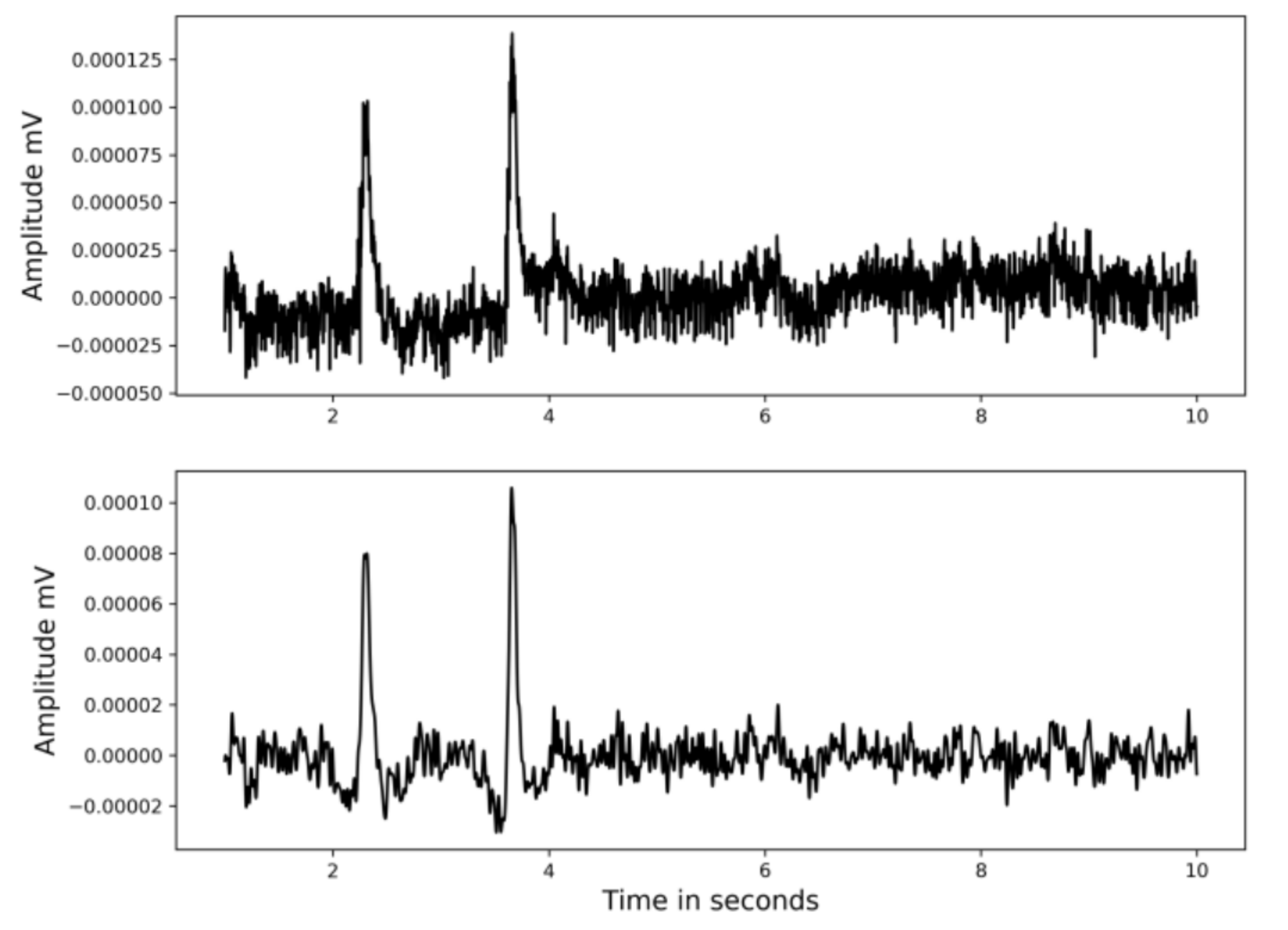}
                \caption{Differences between the raw EEG and filtered data.}
                \label{fig:eeg_example3}
        \end{minipage}%
        \hfill%
        \begin{minipage}[t]{0.48\linewidth}
                \includegraphics[width=\columnwidth,keepaspectratio]{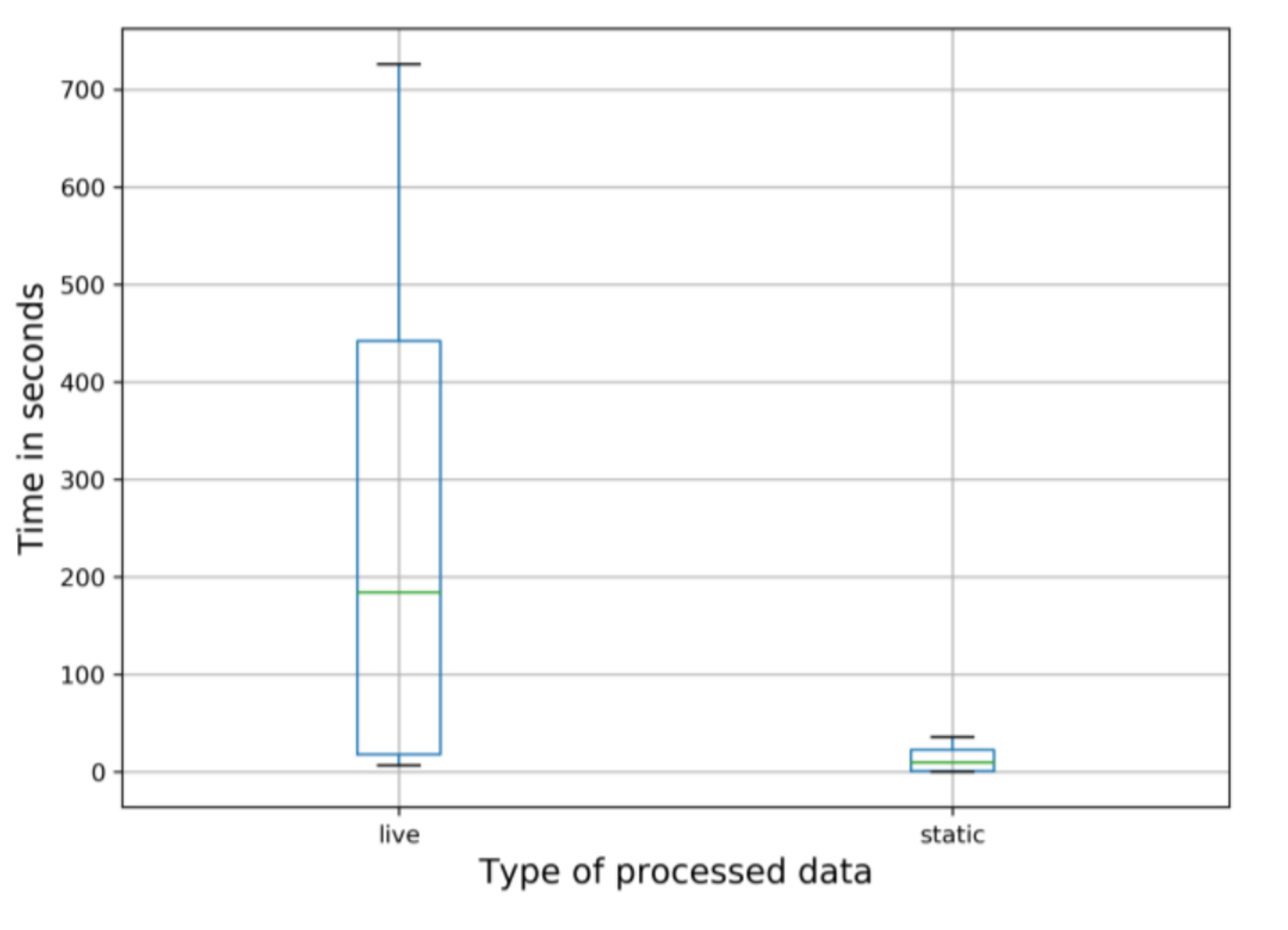}
                \caption{Filtering execution time for the static and live data. The confidence interval is 95\%.}
                \label{fig:eeg_example4}
        \end{minipage}
\end{figure}

We first compare the execution time and accuracy of the applied filtering for the static raw EEG data and stream of the EEG data which simulates a real time brain signal. 
To this end, the full raw EEG data was preprocessed using the band-pass filter that applies to the frequencies from 2 Hz to 30 Hz. As suggested by Widmann et al. \cite{WIDMANN201534}, the band-pass filter uses the minimum frequency of 2Hz to avoid distortions. The first 10 seconds of the raw and filtered EEG data are illustrated in Figure \ref{fig:eeg_example3}. As we can see, the frequencies which are below or above the pass-band are attenuated, producing a filtered signal. Note that we use the desktop-grade system to filter the EEG data in this experiment. 

\subsection{Filtering execution time for static and live data}
We use the cProfile tool to profile the filtering execution time for the static EEG data and live stream of EEG data, which is simulated in our study by splitting brain signals into packets (the size of each packet is 400 data points). Note that in this experiment, we repeat filtering 50 times for each packet and the static data to increase the statistical significance of the measured time.

Figure \ref{fig:eeg_example4} shows the measured execution time of the filtering phase for the static and live EEG data (the confidence interval is 95\%). We see that that when we apply filtering to the static data it takes far less time to process the data compared to the filtering time of the real time data. We attribute this to the fact that the filtering function is invoked to process each packet from the live data. While we invoke this function only once for the static data which is filtered as one packet. Therefore, we may conclude that in case of the live data the filtering time is mostly determined by the time required to execute the processing function.

\subsection{Variation of the filtering accuracy across channels}

The preprocessing phase must provide a correctly filtered EEG signal for both data types, which then can be used to detect certain events. If after filtering, the signal is deteriorated, it can lead to incorrect results and produce wrong control commands for the interface. Thus, we need to ensure that the simulation of real time data acquisition is filtered in the same way, or at least with the harmless amount of distortion, compared to the data filtered statically as one packet. To this end, we use the Pearson correlation coefficient test to compare, channel by channel, both types of filtered EEG signals, the dynamic and static one. 

\begin{figure}[hbt!]
        \begin{minipage}[t]{0.48\linewidth}
                \includegraphics[width=\columnwidth,keepaspectratio]{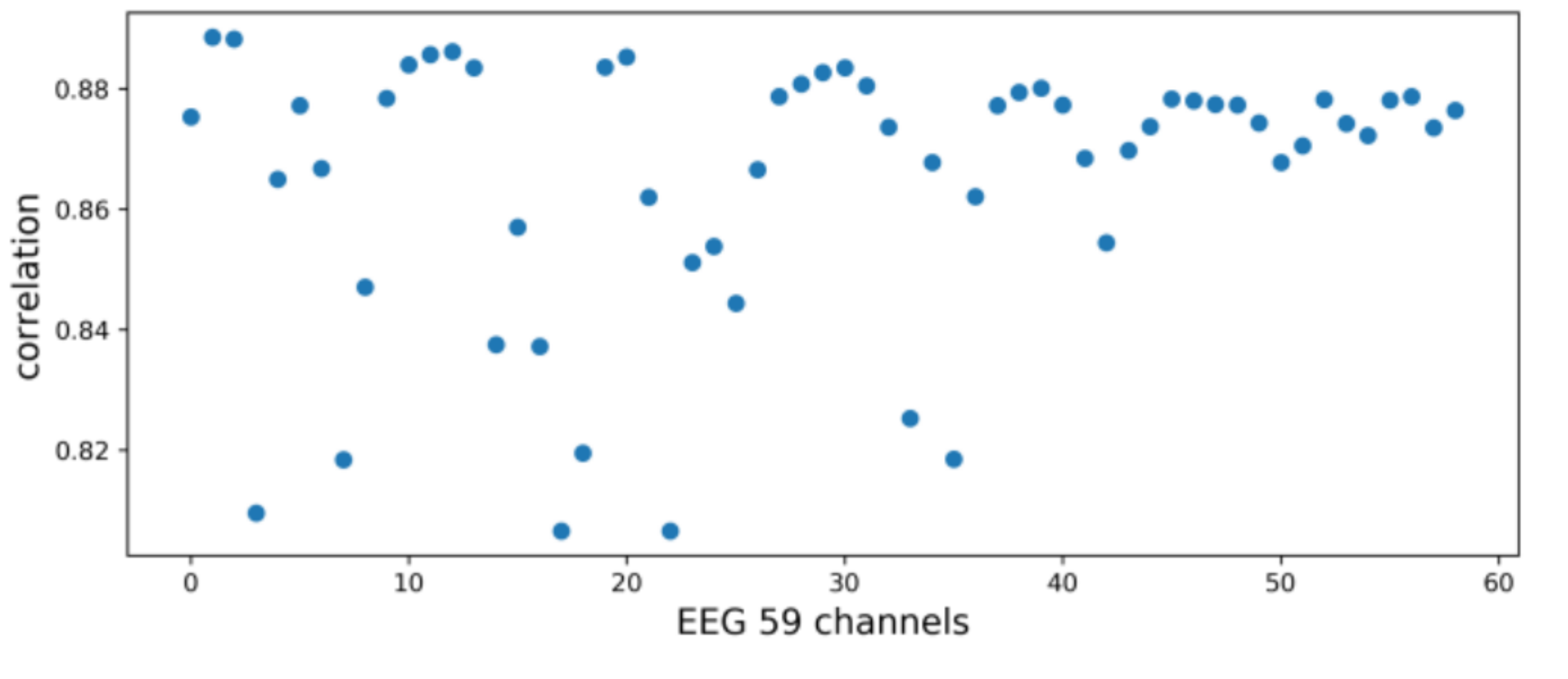}
                \caption{The Pearson coefficient obtained using 59 channels for the filtered static and live data.}
                \label{fig:eeg_example5}
        \end{minipage}%
        \hfill%
        \begin{minipage}[t]{0.48\linewidth}
                \includegraphics[width=\columnwidth,keepaspectratio]{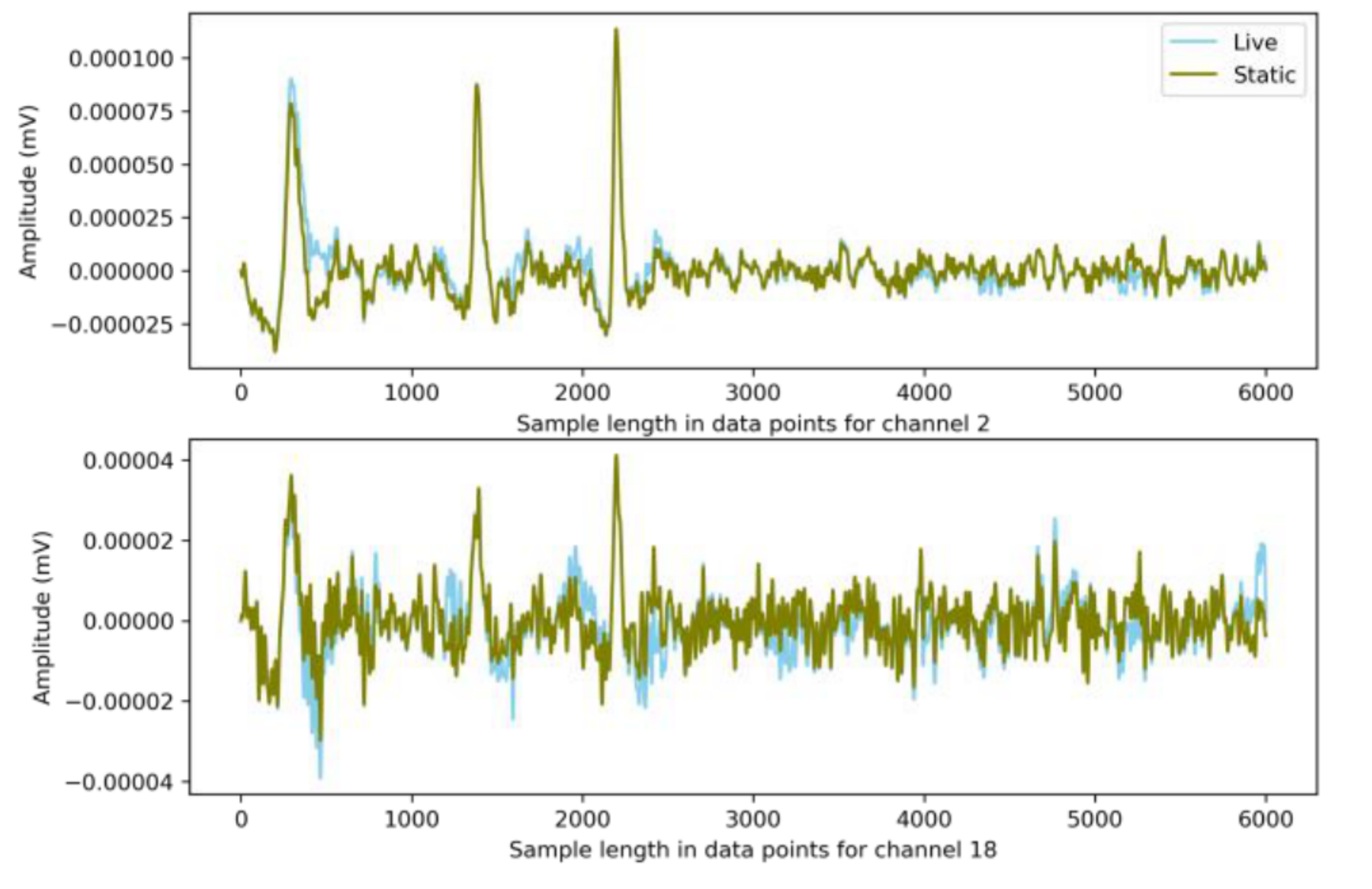}
                \caption{The filtered static and live data for channel 2 and channel 18.}
                \label{fig:eeg_example6}
        \end{minipage}
\end{figure}

Figure \ref{fig:eeg_example5} shows the correlation  coefficient for all the EEG channels. We see that this coefficient fluctuates between 0.81 and 0.88, which indicates that there is a strong correlation between the filtered static and live data. In other words, when the EEG signals are collected in real time with the sampling period of 0.666 seconds (i.e. each packet contains 400 data points) and preprocessed by the band-pass filter, the output does not differ significantly from the data filtered as a whole pre-recording, i.e. without splitting into packets.

Graphs in Figure \ref{fig:eeg_example6} superposes a filtered segment of the EEG signal of both types (static and live) of the processed data for two channels, 2 and 18: the top plot corresponds to channel 2 for which we obtain the highest correlation coefficient, while the bottom plot reflects the graphs for channel 18 where we observe the lowest correlation coefficient. We see that both types of the filtered signals follow the same pattern. Compared to the static data, the filtered live data tends to have a difference in the amplitude at some points, indicating that the band-pass filter parameters and settings have different effect on a brain signal which depends on the length of the signal (or the size of a packet). 

\subsection{The effect of the packet size on filtering accuracy}
To investigate how the packet size affects filtering accuracy, we filtered the brain signal varying the size of packets and compared the results of filtering with the signal which has been filtered as one packet, i.e. the statically filtered signal. Note that in this experiment we also use the Pearson test to compare filtered signals. We varied the size of packets in the following range: 200, 300, 800, 991, 1200 data points. Most of the packet sizes have been chosen to be proportional to the total number of data points in the original signal, i.e. 166800 data points. The only exception is the packet size of 991 data points, the default packet size in the band-pass filter function.

\begin{figure}[!htb]
        \centering
        \includegraphics[height=1.5in,keepaspectratio]{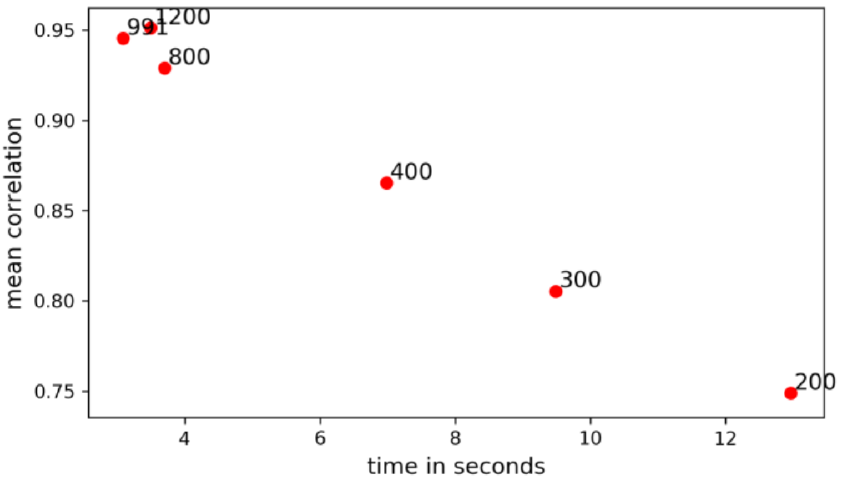}
        \caption{The Pearson test for the filtered signal split into packets and the signal filtered as one packet. Red labels correspond to the size of packets.}
        \label{fig:correlation_accuracy}
\end{figure}

Figure \ref{fig:correlation_accuracy} shows the Pearson test for the filtered signal split into packets and the statically filtered signal. We clearly see that the filtered signals split into packets with the size which exceeds 800 data points have a strong correlation with the statically filtered signal. While the lowest correlation coefficient (~0.75) is observed for the signals split into packets with the smallest size, i.e. 200 data points. Thus, we may conclude that splitting of the signals into packets, which is required for live processing and real time BCI devices, reduces the accuracy of filtering.

\subsection{The effect of the packet size on filtering execution time}
We run filtering with different sampling rates and packet sizes on the desktop-grade machine, as specified in Section 4.2. To find how the size of the samples affects the filtering time, we profiled the filtering stage using cProfile.  Note that we obtain small packets when the signal sampling rate is high, i.e. more packets should be filtered when we increase the sampling rate and reduce the size of packets. For example, the total number of samples is 139 for packets with 1200 data points, while the packets with 200 data points incur 834 samples.
\begin{figure}[hbt!]
        \begin{minipage}[t]{0.48\linewidth}
                \includegraphics[width=\columnwidth,keepaspectratio]{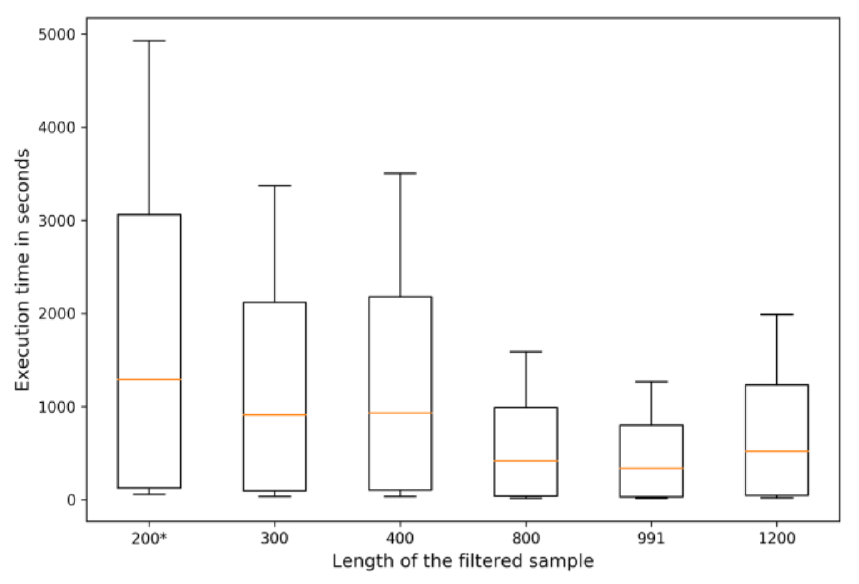}
                \caption{Filtering execution time for packets with 200, 300, 400, 800, 991 and 1200 data points.}
                \label{fig:eeg_execution_time}
        \end{minipage}%
        \hfill%
        \begin{minipage}[t]{0.48\linewidth}
                \includegraphics[width=\columnwidth,keepaspectratio]{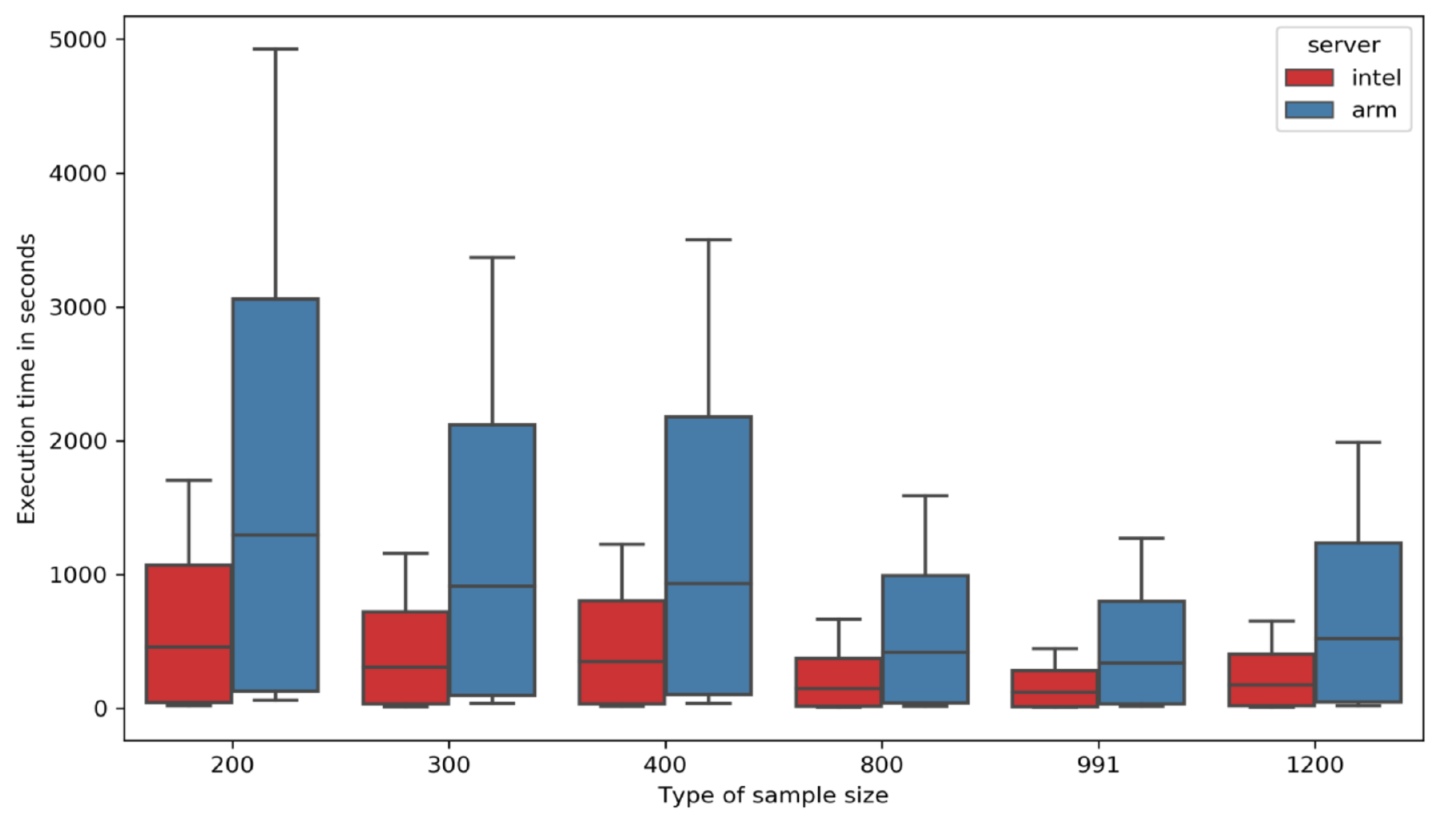}
                \caption{Filtering execution time measured on servers for packets with 200, 300, 400, 800, 991 and 1200 data points.}
                \label{fig:eeg_execution_time_cloudedge}
        \end{minipage}
\end{figure}
Figure \ref{fig:eeg_execution_time} shows the filtering execution time (the confidence interval 95\%) measured for packets with different sizes. Note that we run the filtering function for each packet 100 times to obtain statistically significant measurements. We see that the filtering time decreases with the size of the packets. We explain this by the fact the filtering time of a packet does not grow significantly with the number of data points. Meanwhile, each time we execute the filtering function, we invoke the MNE library function which has a significant delay. As a result, the filtering time of the entire signal is mostly determined by the number of packets which we need to process but not the size of these packets. Nonetheless, we see that the filtering time slightly grows when we increase the number of data points in packets with the size above 991. This implies that for the packets where the total number of data points exceeds 991, the filtering time is determined by the number of data points but not the number of packets. Overall, the packet size of 991 data points provides the minimum filtering time in our experiments. Moreover, according to Figure \ref{fig:correlation_accuracy}, this packet size also provides a high accuracy. Thus, we suggest that this packet size provides the best trade-off between the accuracy and filtering time. Nonetheless, this packet size can be too big for live signal processing and it may not be acceptable for the specific BCIs where the BCI response time should be very small.   

\subsection{Experiments on Cloud and Edge servers}
After performing experiments with the desktop-grade machine, we made a series of experiments on the Intel and Edge servers. Figure \ref{fig:eeg_execution_time_cloudedge} shows how the filtering time changes with the packet size for both Intel and ARM servers. We see that the distribution of filtering times across different packet sizes follow the distribution depicted in Figure \ref{fig:eeg_execution_time}. Thus, we may conclude that  the packets with 991 data points provide the best trade-off between accuracy and filtering time on both Cloud and Edge servers. However, running the filtering phase on the Intel server takes much less time compared to the ARM server. 

\begin{table}
\caption{Performance counters and power measured for the filtering phase running on servers.}
\label{tab1e_counters}
\centering
\begin{tabular}{|l|l|l|}
\hline
 \textbf{Counter} & \textbf{Intel} & \textbf{ARM}\\
\hline
\hline
Cycles & 8,983,280,143 & 16,281,340,001\\
\hline
Instructions & 11,114,539,547 & 10,113,631,460\\
\hline
L1-dcache-loads & 2,963,608,650 & 4,523,218,688\\
\hline
L1-dcache-stores & 1,914,425,122 & 4,523,207,449\\
\hline
LLC-loads & 54,398,598 & 1,418,072,107\\
\hline
LLC-stores & 31,874,277 & 145,184,309\\
\hline
\hline
Power & 118W & 59W \\
\hline
\end{tabular}
\end{table}

Finally, we characterized the single-threaded version of the filtering phase using performance counters for both architectures. Table \ref{tab1e_counters} shows the measurements collected for most significant performance counters, i.e. the number of instructions, cycles, the number of L1 data cache accesses and the number of Last Level Cache (LLC) accesses. In this experiment, we use the packets with the size of 400 data points (1 iteration). Our first observation is that it took almost 2x less cycles for the Intel server to execute the same filtering code which has been executed on the ARM server, even though the Intel server issued more instructions. As a result, the average IPCs are 1.23 and 0.62 for the Intel  and ARM processors, respectively. Thus, as it was expected, we see that the Intel server has a better architectural performance compared to the ARM server. Moreover, we also see that the MNE python library is better optimized for the Intel micro-architecture, since the filtering phase incurs 1.52x more L1 data load instructions and 2.36x more L2 data store instructions when running on the ARM processor. As a result, the ARM processor issue 26x more LLC load instructions and 4.7x more LLC store instructions. Nonetheless, as we see in Table \ref{tab1e_counters}, the ARM server consumes almost 2x less power than the Intel server.

To sum up, the Intel server filters the brain signal much quicker, up to 3.5x, than the ARM server. However, the Cloud facilities might not provide a quick reply in case of implementing a real time BCI device due to a high network latency. Meanwhile, the Edge facilities have a much lower network latency, but as we see a typical Edge server has a lower performance compared to the Cloud Intel server, even though it has better power efficiency.
In our future research, we will investigate particular examples of BCI devices and how efficient different Edge and Cloud architectures  and networks for running the entire BCI pipeline, including the filtering phase. 
\section{Conclusion}
In this paper, we present the results of our preliminary study on the accuracy and performance of a Brain-Computer Interface performing on typical Cloud and Edge servers. We demonstrate how the accuracy and performance of the pre-processing phase, i.e. the brain signal filtering phase, differs when processing static and live streaming data, obtained in real time BCI devices. We identify the optimal size of packets for sampling the brain signal in real time.  Finally, we demonstrate efficiency of using typical Cloud and Edge servers for running the BCI filtering phase.

\bibliographystyle{IEEEtran}
\bibliography{IEEEabrv,main}
\end{document}